\documentclass[12pt]{article}
\usepackage{graphicx}

\newcommand\pubnumber{}
\newcommand\pubdate{\today}

\def\napoli{Ruhr-Universit\"at Bochum, Experimentelle Hadronenphysik\\
44780 Bochum, Germany
}

\def\Title#1{\begin{center} {\Large #1 } \end{center}}
\def\Author#1{\begin{center}{ \sc #1} \end{center}}
\def\Address#1{\begin{center}{ \it #1} \end{center}}

\newcommand\pubblock{\rightline{\begin{tabular}{l} \pubnumber\\
         \pubdate  \end{tabular}}}
\newenvironment{Abstract}{\begin{quotation}  }{\end{quotation}}
\newenvironment{Presented}{\begin{quotation} \begin{center} 
             PRESENTED AT\end{center}\bigskip 
      \begin{center}\begin{large}}{\end{large}\end{center} \end{quotation}}

\def\beq{\begin{equation}}
\def\eeq#1{\label{#1}\end{equation}}
\def\eeqn{\end{equation}}

\def\beqa{\begin{eqnarray}}
\def\eeqa#1{\label{#1}\end{eqnarray}}
\def\eeqan{\end{eqnarray}}

\let\bar=\overbar

\def\Dslash{\not{\hbox{\kern-4pt $D$}}}
\def\dslash{\not{\hbox{\kern-2pt $\del$}}}

\def\BR{\mbox{\rm BR}}

\def\msb{{\bar{\ssstyle M \kern -1pt S}}}

\input babarsym.tex

\begin{document}
\begin{titlepage}
\pubblock

\vfill
\Title{Prospects for \panda in Charmonium and Charm Physics}
\vfill
\Author{ Marc Peliz\"aus\\(for the \panda Collaboration)}
\Address{\napoli}
\vfill
\begin{Abstract}
The prospects of the future PANDA experiment at FAIR in Darmstadt/Germany in the field of 
charmonium and charm spectroscopy are discussed.
\end{Abstract}
\vfill
\begin{Presented}
Charm2012\\
Honolulu, Hawaii, USA, May 14-17
\end{Presented}
\vfill
\end{titlepage}
\def\thefootnote{\fnsymbol{footnote}}
\setcounter{footnote}{0}

\section{The \panda Experiment and Physics Program}
The \panda experiment \cite{Lutz:2009ff} to be build at the future
site FAIR (Facility for Antiproton and Ion Research) in Darmstadt in
Germany is optimized for high precision hadron physics in the
charmonium mass region. The experiment utilizes an antiproton beam
with high precision and high luminosity as well as a versatile
detector. The produced antiprotons at FAIR are stored and cooled in
the HESR (High Energy Storage Ring) with a momentum resolution of up
to $10^{-5}$ in the momentum range between 1.5 to 15\gevc. The
antiprotons can collide with the internal target of the \panda
experiment, providing a luminosity of up to
$10^{32}\cm^{-2}s^{-1}$. The detector consists of a target
spectrometer with a solenoid magnet surrounding the interaction region
and a forward spectrometer with a dipole magnet. The detector covers
almost $4\pi$ of the solid angle. Precise vertex reconstruction is
achieved by a vertex detector in the target spectrometer. Charged
particle tracking with high resolution, precise electromagnetic
calorimetry over a wide energy range as well as muon identification
are provided by both spectrometers. DIRC detectors in the target
spectrometer complete the particle identification system.

The experimental setup allows to address many topics in hadron
physics, including the study of QCD bound states, non-perturbative QCD
dynamics, hadrons in nuclear matter, hyper-nuclear physics,
electromagnetic processes as well as electroweak physics. The
addressed topics are discussed in detail in \cite{Lutz:2009ff}. In this
contribution the prospects of \panda in charmonium and open charm
spectroscopy are highlighted.

\section{Spectroscopy in the Charm Mass Region}
\subsection{Charmonium System}
Charmonium spectroscopy has been proven a very powerful tool for the
understanding of the strong interaction in the past decades.  Due to
the high mass of the charm quark, the \ccbar system can be treated in
non-relativistic potential models, where the potential is chosen to
reproduce the properties of the strong interaction and the free
parameters are obtained from comparison to data. While the gross
features of the spectrum are reasonable described by these models,
calculations in various theoretical frameworks are carried out to
understand the finer features\cite{Brambilla:2010cs}. To distinguish between the various approaches, 
precise experimental data is required.

All eight predicted charmonium states below the open charm
threshold have been identified experimentally\cite{Beringer:2012}. Above the threshold only a few of the
predicted states have been classified. On the other hand many states
referred to as the X,Y, and Z states have been discovered recently in
this mass region. The unexpected properties of some of these states
make it difficult to interpret these as conventional charmonium
states and their nature is controversially discussed \cite{Pakhlova:2010zza}. Resolving the
situation above the threshold is an important task in spectroscopy.

\subsection{Gluonic Excitations}
For a complete understanding of the hadron spectrum it is important to
take the coherent interaction of the gluons into account , which can
manifest in the formation of a gluon tube between the constituent
quarks. Excitations of this tube will add additional degrees of
freedom to the hadronic bound state. This form of hadrons is referred
to as hybrids, which will appear as supernumerary to the \qqbar states
expected from the naive quark model. In the simplest scenario the
\ccbar bound state is in an $S$-wave configuration and the flux tube in
a $1^-$ color-magnetic or $1^+$ color-electric excitation resulting in
the lightest 8 hybrid states, where three states have exotic quantum
numbers $J^{PC}$, which are forbidden for a \qqbar system.  Most
predictions agree that the $1^{-+}$ state is the lightest charmonium
hybrid with exotic quantum numbers having a mass of about 4300\mevcc
\cite{Chen:2001}.  Predicted decay modes for charmonium hybrids are
$\chic{1}\piz\piz$, $\jpsi\piz\piz$, $\jpsi\omega$ and $D\Dstarb$. In
the light quark sector the $\pi_1(1400)$ and $\pi_1(1600)$ with
$J^{PC}=1^{-+}$ are discussed as hybrids. Both are also observed in
\pbarp annihilation at rest with rates comparable to those of
conventional mesons. So there is a large potential for \panda to
discover such states in the charmonium sector.

It is also expected that the self-interaction of gluons can form
glueballs, bound states consisting of only gluons. Lattice QCD
calculations predict the glueball ground state to be a scalar with a
mass of about 1500\mevcc \cite{Morningstar:1999}. The $f_0(1500)$ observed in \pbarp
annihilation is discussed to be a glueball. Excited glueballs are
predicted to fall into the mass range of the charmonium system. In
particular the lightest states with exotic quantum numbers are
predicted at 4100\mevcc ($2^{+-}$) and 4740\mevcc ($0^{+-}$). Since
gluons do not distinguish between flavors, glueballs should couple to
hadronic final states independently of their flavor (flavor
blindness). Thus heavy glueballs should, like charmonia and hybrids,
also decay into hidden charm states and $D$ meson pairs.

\subsection{Open Charm Mesons}
Containing a heavy and a light quark, $D$ mesons are very interesting
objects for the understanding of QCD. In the heavy quark limit
$m_c\to\infty$, the charm quark can be seen as a static color source.
On the other hand, the light quark introduces chiral symmetry breaking
and restoration. Until 2003 the spectrum of $D_s$ mesons was regarded
as well understood in potential model
calculations\cite{Isgur:1991}. At that time the two narrow states
$D_{s0}^*(2317)^+$ and $D_{s1}(2460)^+$ have been discovered. The
masses of these states are considerably lower than predicted. To
explain this discrepancy a series of theoretical work was carried out
in various frameworks\cite{Cahn:2003}. Interpretations include
conventional $c\sbar$ states, tetra-quark states and -- for
$D_{s0}^*(2317)^+$ -- a $DK$ molecule.

\section{Prospects for \panda}
\subsection{Mass and Width Scans}
In antiproton-proton annihilation all states with non-exotic quantum
numbers $J^{PC}$ can be accessed directly in formation. This allows
\panda to carry out scan experiments, where the cross section of a
resonance is measured as a function of the center of mass energy in
fine steps. From the observed cross section one can measure the mass
and width of the scanned resonance very precisely, since the
measurement is not limited by the detector resolution. The
uncertainties are rather given by the beam momentum, which is known
very precisely. The method has been proven very
successful by the Fermilab \pbarp annihilation experiments E760 and
E835. The E835 measurement \cite{Andreotti:2005ts} of the \chic{1} mass and widths
($m=(3510.719\pm 0.051\pm0.019)\mevcc$, $\Gamma=(876\pm45\pm26)\kev$)
are the most precise up to date.

Having direct access to all charmonium states with masses below
5\gevcc and performing measurements with highest precision will put
the \panda experiment in a unique position in the next decade.  In
particular the $\eta_c$ and $\eta_c(2S)$ masses can be measured to
precisely obtain the $1S$ and $2S$ hyper-fine mass splitting. Likewise the
precise determination of the $h_c$ mass and its position with respect
to the spin-triplet \chic{} states will provide more insight into the
spin-dependence of the long-range potential. A detailed simulation for
the measurement of the $h_c$ reconstructed in its radiative decay to
\etac, followed by $\etac\to\phi\phi$ and $\phi\to\Kp\Km$ was
performed and sufficient background rejection and efficiency is
obtained to allow a scan with 10 scan points and 40 days of data
taking\cite{Lutz:2009ff}.

Among the newly discovered states above the open charm threshold is
the $X(3872)$, a narrow state ($\Gamma<1.2\mev$) with $J^{PC}=1^{++}$
or $2^{-+}$ decaying into $\jpsi\pi\pi$ and
$D\Dstarb$\cite{Beringer:2012}. The mass of the state is
$(3871.68\pm0.17)\mevcc$. It is widely believed that the closeness of
the $X(3872)$ to the $D\Dstarb$ mass threshold is related to its
nature. It is suggested that the $X(3872)$ is a weakly bound
$D\Dstarb$ molecule below or a virtual state above the
threshold\cite{Hanhart:2009}. The current experimental uncertainties
do not allow to distinguish between the two scenarios. It is also
proposed, that the line shape of the $X(3872)$ in the $\jpsi\pi\pi$
and $D\Dbar\piz$ decay modes might allow to distinguish between the
two possibilities. A precise scan of the resonance in both decay modes
at \panda will improve the experimental situation on the $X(3872)$
dramatically\cite{Lange:2010}.

\subsection{Exotic Charmonium Hybrid}
The lightest exotic charmonium hybrid is predicted to be a $1^{-+}$
state in the mass region of about 4300\mevcc. Its width can be narrow
and decays to open and hidden charm are expected. Two prime decay
modes could be $\chic{1}\piz\piz$ and $D\Dstarb$. Any state with
exotic $J^{PC}$ can only be accessed in production with an associated
recoil system. In a detailed simulation study the production of the
exotic hybrid state with a recoiling $\eta$ has been investigated at a
center-of-mass energy of $\sqrt{s}=5.38\gevcc$.  The hidden charm
intermediate state is reconstructed from the $\chic1\to\jpsi\gamma$
and $\jpsi\to\epem$ decays, leading to a final state with two leptons
and seven photons. The open charm decay mode is reconstructed from the
$\Dstar\to D\piz$ and $D\to\Kp\Km\piz$ decay modes, leading to a final
state with four charged kaons and eight photons. Reactions with only
light hadrons have been proven to be suppressed sufficiently for
$D\Dbar$ and $\Dstar\Dstarb$ production. Other backgrounds are
reactions with an event topology similar to that of the signal
reaction like $\pbarp\to\chic{1}\pi\eta\eta$,
$\chic{1}\piz\piz\piz\eta$, $\jpsi\piz\piz\piz\eta$, $D\Dbar\piz$.
Excellent calorimetry is required to suppress these backgrounds. From
the study a signal to background ratio of better than 10 is expected,
if the cross sections for these reactions do not exceed the signal
cross section by more than an order of magnitude.

Since no assumptions are made on the particular nature of the produced
state in the study, the results are valid for any narrow object of a
mass of about 4300\mevcc. In particular the investigated $D\Dstarb$
decay would be interesting for heavy glueball searches.

\subsection{Charged Charmonium-like States}
Among the new charmonium-like resonances are three charged states
$Z(4430)^+$, \linebreak $Z_1(4050)^+$ and $Z_2(4250)^+$, discovered by Belle in
$B\to ZK$ decays. The $Z(4430)^+$ is observed in its decay into
$\psitwos\pip$ and $Z_1(4050)^+$ and $Z_2(4250)^+$) in their decays
into $\chic{1}\pip$ \cite{Choi:2008}. Since they decay into a
charmonium state and carry electric charge, their minimal quark
content must be $\ccbar u\dbar$ and thus they are definitely of exotic
nature. Confirmation from another experiment is still missing. All
three states have been searched for at Babar\cite{Aubert:2009}. No
evidence was reported, but the upper limits set by Babar do not
strictly rule out Belle's measurements.

At \panda these states can be studied in \pbarp production, in a
simplest scenario together with a recoiling \pim. A further option for
\panda is to study these states in formation on a deuterium target
($\bar{p}d\to Z^- p$) with a spectator proton.

\subsection{Open Charm Spectroscopy}
An important quantity possibly allowing to distinguish between the
different theoretical explanations to the $D_{s0}^*(2317)^+$ and
$D_{s1}(2460)^+$ \cite{Cahn:2003} is the decay width of these
states. Currently the widths are constrained by upper limits of about
3\mev due to the experimental resolution. This is not sufficient to
distinguish between the different theoretical approaches and draw
further conclusions about the internal structure of these states.

At \panda the $D_{sJ}^+$ mesons, $D_{s0}^*(2317)^+$ and $D_{s1}(2460)^+$,
can be produced in \pbarp annihilation with a recoiling $D_s^-$ meson.  The
cross section for the process $\pbarp\to D_{sJ} D_s^-$ depends on the
width of the $D_{sJ}^+$. By measuring the cross section in dependence of
the center-of-mass energy near the threshold in fine energy steps, the
$D_{sJ}^+$ width can be extracted from the observed cross section. This
method is expected to be sensitive down to widths of about 100\kev \cite{Lutz:2009ff}.

\end{document}